% TeX file
\magnification=\magstep1
\centerline{Lambda Oscillations and the Conservation Laws}
\medskip
\centerline{A. Widom and Y.N. Srivastava}
\centerline{Physics Department, Northeastern University, Boston MA, USA}
\medskip
\par
\noindent
{\it ABSTRACT:} Lowe, Bassalleck, Burkhardt, Rusek, Stephenson, 
and Goldman assert (under the assumption that secondary decay vertices 
exhibit amplitude interference at fixed space-time points) that 
for $\pi^-+p^+\to \Lambda +K^0$  Lambda oscillations disappear. 
Under the same assumption, we find that conservation of energy and 
conservation of momentum also disappear. Quantum oscillations occur 
for quite ordinary particles for reasons which are discussed.  
\bigskip
\centerline{\bf 1: Introduction}
\medskip

In two particle ``in'' and two particle ``out'' scattering, for which
$$
\pi^-+p^+\to \Lambda +K^0, \eqno(1a)
$$
is a special case, there is a well known theorem that the spatial wave 
functions (for fixed total energy and fixed total momentum) factor into 
a product of wave functions; one for the ``center of mass coordinate'' 
${\bf R}$, and one for the ``relative coordinate'' ${\bf r}$, 
$$
\Psi_{total}({\bf r}_1,{\bf r}_2)=\phi (\bf R)\psi({\bf r}).    \eqno(1b)
$$
Discrete quantum numbers are left implicit. While the wave function 
$\phi ({\bf R})$ is in principle a ``wave packet'', it is still 
conventional to use a plane wave in a very big box and write 
$\phi ({\bf R})=exp\{ i({\bf P}_{total}{\bf \cdot R})/\hbar\}$. It is 
also usual to work in the center of mass frame in which case 
${\bf P}_{total}={\bf 0}$. This convention will be followed below. 

Hence, in the center of mass frame ${\bf P}_{total}=0$ (and for the 
moment neglecting life time effects), the conservation laws of energy 
and momentum dictate for Eqs.(1) that the ``in'' wave function is 
given by  
$$
\Psi_{in}({\bf r}_p,{\bf r}_\pi)=\psi_{in}({\bf r}_p-{\bf r}_\pi). 
\eqno(2a)
$$ 
and that the ``out'' wave function is given by 
$$
\Psi_{out}({\bf r}_K,{\bf r}_\Lambda)=
\psi_{out}({\bf r}_K-{\bf r}_\Lambda). \eqno(2b)
$$
With Eq.(2b) in mind, we point out the following
\medskip
\par \noindent
{\it Theorem:} If a wave function 
$\psi_{out}({\bf r}_K-{\bf r}_\Lambda)$ 
oscillates in the coordinate ${\bf r}_K$, then the same 
wave function oscillates in the coordinate ${\bf r}_\Lambda$
\medskip
\par \noindent
We trust that the proof is obvious. If the $K^0$ oscillates in space, 
then the $\Lambda $ oscillates in space. And this will be reflected in 
the positions of the secondary vertices of the joint distribution of both 
particles via $|\Psi_{out}({\bf r}_K,{\bf r}_\Lambda)|^2$. 

With life-time effects {\it included}, we discussed$^{[1]}$ $\Lambda $ 
oscillations in previous work. More recently, Lowe, Bassalleck, Burkhardt, 
Rusek, Stephenson, and Goldman$^{[2]}$ find that the Lambda oscillations 
disappear, but ``normal'' Kaon oscillations are left in tact. To get 
to the root cause of the differences in opinion, it is then simply a matter 
of locating the point at which momentum and energy are no longer conserved 
in the argument of Lowe et. al., and this we shall do in what follows.
\medskip
\centerline{\bf 2: Momentum Conservation}
\medskip
Shown schematically below is the outgoing state of Eqs.(1)
$$
\matrix
{\ &{\bf p}_\Lambda &\ & {\bf p}_K &\cr
\Lambda &\leftarrow &(out)&\rightarrow & K^0}. \eqno(3)  
$$
Note that in the center of mass frame 
$$
{\bf P}_{total}={\bf p}_\Lambda +{\bf p}_K={\bf 0}, \eqno(4a)
$$
so that 
$$
exp\{i({\bf p}_\Lambda {\bf \cdot r}_\Lambda +
{\bf p}_K {\bf \cdot r}_K)/\hbar \}=exp\{i{\bf p}_K {\bf \cdot}
({\bf r}_K -{\bf r}_\Lambda )/\hbar \}. \eqno(4b)
$$
In decomposing the process in Eq.(3) into various possibilities, and 
then superimposing amplitudes for these possibilities, we enforce strict 
conservation of momentum, which means in the center of mass system 
that Eqs.(4) must be enforced. The ``long'' and the ``short'' of it 
are shown below 
$$
\pmatrix
{\ &-{\bf p}_L &\ & {\bf p}_L &\cr
\Lambda &\leftarrow &(out)&\rightarrow & K_L} and 
\pmatrix
{\ &-{\bf p}_S &\ & {\bf p}_S &\cr
\Lambda &\leftarrow &(out)&\rightarrow & K_S}, \eqno(5) 
$$
and the $\Lambda $ momentum is always equal and opposite to the $K$ 
momentum be it a long $K_L$ with mass $M_L$ and momentum ${\bf p}_L$ 
or a short $K_S$ with a mass $M_S$ and momentum ${\bf p}_S$. In any 
case, from momentum conservation laws in Eqs.(4), superpositions 
of all the processes yield the form of the wave 
function $\psi_{out}({\bf r}_K-{\bf r}_\Lambda)$. It is 
simply impossible to construct a wave function of that form which 
oscillates in the coordinate ${\bf r}_K$ but does not oscillate in 
the coordinate ${\bf r}_\Lambda$.
\medskip 
\centerline{\bf 3: Proper Times}
\medskip 

The violation of the four momentum conservation law is slipped into 
the argument of Lowe et. al. by considerations of ``proper time''. Here 
is the story of proper time:

Once upon a time there were two twins who led different life 
styles. One twin (along with laboratory observers) hardly moved at all, 
while the other twin moved around pretty fast. Much to the shock of the 
laboratory observers, the fast twin seemed younger than the slow twin. 
The fast twin's proper time was never the same as the 
laboratory observers' proper time, and it was not even the same as the 
proper time that professor Einstein taught the laboratory observers to 
calculate
$$
c^2\tau^2=c^2t^2-|{\bf r}|^2. \eqno(6)
$$
Excuse us for telling the wrong (classical) story. Einstein already 
explained to all of us that the two twins read out two different 
proper times so there is no paradox! 

Please let us start again. Once upon a time one electron was fired 
at two slits, so the electron split into two virtual twins who later met 
and combined again at the same counter behind the slits. One virtual twin 
went through slit 1 and arrived at the counter in her proper time $\tau_1$. 
The other virtual twin went through slit 2 and arrived at the same 
counter with her proper time $\tau_2$. Nevertheless, it was really only one 
electron but with two proper times and Einstein never liked this story. 
Bohr told Einstein that there was no paradox! The laboratory observers 
where shocked to learn that the virtual twin electrons had the same 
energy, traveled different distances to the same counter, but arrived at 
this counter at the same laboratory time. Bohr told the laboratory 
observers to {\it stop thinking in classical terms}! The two paths, and two 
proper times, and the diffraction oscillation from a beam of such electrons 
in the counters behind the slits, were all simply a consequence of the wave 
function and the interference phase 
$\theta_{12}=(mc^2/\hbar )(\tau_2-\tau_1)$ where $m$ is the electron mass.  
If one bounces electrons off a crystal, then reflected out 
going waves oscillate. Low energy electron diffraction (LEED) is measured 
every day. Ordinary particles can oscillate!  
 
Now in high energy physics, one measures a vertex position ${\bf r}$ and 
a momentum ${\bf p}$ and deduces for a particle of mass $M$ what is often 
called proper time, 
$$
\tau_{experimental}=(M|\bf r|/|{\bf p}|). \eqno(7) 
$$
The important point is that Eq.(7) would be true if the particle were 
classical. But what if the particle is quantum mechanical? It would be 
hard to maintain from the viewpoint of the uncertainty principle that 
{\it both} ${\bf p}$ and ${\bf r}$ are exactly known or that {\it both} 
$E$ and $t$ are exactly known. Furthermore, the laboratory observers do 
not even attempt measure the time $t$ in Eq.(6). They measure a 
``proper time'' as defined in Eq.(7). 

Consider the $K^0$ particle in Eqs.(1). This meson has ``two proper 
times'' just like the electron going through two slits; i.e. 
$$
\tau_L=(M_L|{\bf r}_K|/|{\bf p}_L|),
\ \ \tau_S=(M_S|{\bf r}_K|/|{\bf p}_S|), \eqno(8)  
$$  
and the laboratory observers might like to know that neither of these 
proper times need be the classical proper time in Eq.(6) which is not 
measured anyway. Nobody even {\it tries} to measure $t$ in Eq.(6). 
If you conserve total energy for both the ``long'' and the ``short'' 
alternatives,  
$$
\sqrt{c^2 |{\bf p}_L|^2+M_\Lambda ^2 c^4}
+\sqrt{c^2 |{\bf p}_L|^2+M_L^2 c^4}
=\sqrt{c^2 |{\bf p}_S|^2+M_\Lambda ^2 c^4}+
\sqrt{c^2 |{\bf p}_S|^2+M_S^2 c^4}, \eqno(9)
$$
then it is evident that you can have neither $M_L$ the same as 
$M_S$ nor $|{\bf p}_L|$ the same as $|{\bf p}_S|$.

For the interference phase of the $K^0$, 
$$
\theta_K=c^2(M_L \tau_L-M_S \tau_S)/\hbar , \eqno(10)
$$
let us define 
$$
\bar{\tau}_K=(1/2)(\tau_L+\tau_S), \ \ \bar{M}=(1/2)(M_L+M_S), \eqno(11a)
$$ 
$$
\Delta M_K=(M_L-M_S), \ \ \Delta{\tau }_K=(\tau_L-\tau_S). \eqno(11b)
$$
Then the two terms in Eq.(10) yield 
$$
\hbar \theta_K=c^2(\bar{\tau}_K \Delta M_K
+\bar{M} \Delta{\tau }_K), \eqno(12)   
$$
which {\it depend on total energy}. Lowe et. al. do not find that 
$\theta_K$ depends on total energy for the very good reason that they 
do not conserve total energy and they do not maintain the two proper times  
$\tau_L$ and $\tau_S$ for the Kaon. One cannot conserve total energy and 
find only one proper time for the oscillating $K^0$. Elimination of one 
of the two proper times also eliminates conservation of energy.

The $\Lambda $ in Eqs.(1) oscillates, but not because the mass is split. 
Like an oscillating diffracted electron wave bouncing off a crystal in 
LEED experiments, the interference phase for the $\Lambda $ has only the 
proper time differences 
$$
\hbar \theta_\Lambda=c^2M_\Lambda \Delta \tau_\Lambda
=c^2 M_\Lambda ^2 |{\bf r}_\Lambda |
\{(1/|{\bf p}_S|)-(1/|{\bf p}_L|) \}. \eqno(13)
$$
Conservation of energy and momentum demands that $|{\bf p}_L|$ 
be different than $|{\bf p}_S|$ so the Lambda has a non-zero 
interference phase $\theta_\Lambda $. This too depends on total 
energy if one conserves total energy.
\medskip
\centerline{\bf 4: Conclusion}
\medskip
It is perhaps unfortunate that quantum ``oscillations'' in high 
energy physics seems to refer to some esoteric particles rather 
than plain quantum interference or diffraction which can occur for any 
quantum particles. One might say, ``the Kaon is very strange; see 
how it oscillates?''. But the electron is not very strange, and it 
oscillates too! The reason that quantum interference occurs 
is that quantum particles do not move on well defined space-time  
paths. For the problem at hand conservation of energy and 
conservation of momentum dictate an internal wave function 
such that the joint probability 
$|\Psi_{out} ({\bf r}_K,{\bf r}_\Lambda |^2$ oscillates in both 
coordinates.
\medskip
\centerline{\bf Acknowledgements}
\medskip
\par \noindent
A. Widom is grateful to professor J. Lowe for many communications. 
\medskip
\centerline{\bf References}
\medskip
\par \noindent
[1] Y.N. Srivastava, A. Widom and E. Sassaroli, {\it Phys. Lett.}
{\bf B334}, 436 (1995).
\par \noindent
[2] J. Lowe, B. Bassalleck, H. Burkhardt, A. Rusek, G.J. Stephenson Jr., 
and T. Goldman, ``No Lambda Oscillations'' hep-ph/9605234.

\bye